\documentclass[graybox,natbib,nosecnum]{svmult}
\bibpunct{(}{)}{;}{a}{}{,} 

\pdfoutput=1   

\usepackage{mathptmx}       
\usepackage{helvet}         
\usepackage{courier}        
\usepackage{type1cm}        

\usepackage{makeidx}         
\usepackage{graphicx}        
\usepackage{multicol}        
\usepackage[bottom]{footmisc}
\usepackage[normalem]{ulem}	
\usepackage{hyperref}  
\usepackage{amssymb}
\usepackage{amsmath}
\newcommand{\thisstar}{WD 1145+017}
\newcommand{\kms}{\,km\,s$^{-1}$}

\newcommand{\Kepler}{{\em Kepler}}
\newcommand{\Spitzer}{{\em Spitzer}}

\usepackage{soul}   

\newcommand*\aap{A\&A}

\newcommand*\aj{AJ}

\newcommand*\apj{ApJ}
\newcommand*\apjl{ApJ}

\newcommand*\apjs{ApJS}

\newcommand*\icarus{Icarus}

\newcommand*\mnras{MNRAS}

\newcommand*\nat{Nature}

\usepackage{color}

\newcommand{\ron}{\color{black}}

\makeindex             

\begin{document}

\title*{Transiting Disintegrating Planetary Debris around WD 1145+017}
\author{Andrew Vanderburg and Saul A. Rappaport}
\institute{Andrew Vanderburg \at Harvard-Smithsonian Center for Astrophysics, Cambridge, MA 02138 \\ 
Current address: Department of Astronomy, The University of Texas at Austin, Austin, TX 78712, USA\\ \email{avanderburg@utexas.edu}
\and Saul A. Rappaport \at Massachusetts Institute of Technology, Cambridge, MA 02139, \email{sar@mit.edu}}
%
%
\maketitle

\abstract{More than a decade after astronomers realized that disrupted planetary material likely pollutes the surfaces of many white dwarf stars, the discovery of transiting debris orbiting the white dwarf WD 1145+017 has opened the door to new explorations of this process. We describe the observational evidence for transiting planetary material and the current theoretical understanding (and in some cases lack thereof) of the phenomenon.}

\section{Overview}
A bit more than a decade after astronomers first began to suspect that white dwarf stars occasionally disrupt and accrete asteroids and small planets from their primordial planetary systems \citep{Debes02, Jura03}, an impressive body of evidence had emerged in support of this scenario. In early 2015, it was known that (a) a large fraction of white dwarf stars (between 25\% and 50\%) are ``polluted'' with trace amounts of elements like silicon, iron, calcium and magnesium in their atmospheres \citep{zuckerman10, koester14}, (b) many of these polluted white dwarfs also showed evidence of warm rocky material orbiting the star in a debris disk \citep{Barber12}, and (c) the abundance ratios of heavy elements in the atmospheres of white dwarfs very closely matched the abundance patterns in rocky bodies in the solar system \citep{Zuckerman06, Farihi13}.

The generally accepted explanation for these observations was that these polluted white dwarfs host planetary systems which at least partially survived the white dwarf progenitor's evolution off the main sequence. The host star's evolution was not without ill effects: as the host shed its outer layers and began to cool and contract into a white dwarf, the star's mass loss caused changes to the planetary system's dynamics. Numerical simulations have shown that planetary systems whose host stars have undergone this type of mass loss can occasionally perturb small planets or asteroids into highly eccentric orbits which can occasionally have periastron passages close enough to the host star (by this time a white dwarf) to be tidally disrupted. The {\ron planetary, lunar \citep{Payne16}, or asteroidal} remnants would then be pulverized into a fine dust (causing the infrared excesses observed around many polluted white dwarf stars) and slowly accreted onto the white dwarf's surface, where the constituent elements would manifest themselves by the presence of spectral lines. 

However, the evidence for this scenario was entirely circumstantial, relying on analysis of the after-effects of planetary disruption. There were occasional detections of transient events indicating possible tidal disruptions in progress \citep{DelSanto14, Xu14}, but no unambiguous detections of disintegrating rocky material until the {\ron discovery} of transits around a polluted white dwarf called \thisstar. For the first time, astronomers had definitively observed the transient process of a white dwarf tidally disrupting a {\ron large rocky body (planet, moon, or asteroid)}, in real time. \thisstar\ has provided the strongest evidence yet that white dwarfs disrupt their planetary systems, and has given astronomers new ways to study and constrain models of the process.

\section{Discovery Observations}
WD 1145-017 was first identified as a somewhat anonymous white dwarf in 1991 by \citet{berg92}, who used low-resolution spectroscopy to classify it as a helium-envelope white dwarf. This classification was confirmed by \citet{friedrich}, although neither groups' classification spectra were strong enough nor taken at high enough spectral resolution to detect any features besides the deep and broad helium lines. Little attention was given to WD 1145+017 until 2014, when it fell in one of the fields of view of NASA's {\em Kepler} Observatory in its extended K2 mission.  During the interval between June and August 2014, K2 observed the field containing \thisstar\ (along with ~20,000 other stars) for a bit less than 80 days. 

Once the K2 data were publicly released, \citet{v15} searched through the data looking for planetary transits. The majority of the stars observed by K2 were main-sequence stars, typically the mass of the sun or {\ron lower}.  However, a handful of other objects, including about 150 white dwarf stars, were proposed by various groups and were observed by K2 as well. 

\begin{figure}
\centering
\includegraphics[scale=.60]{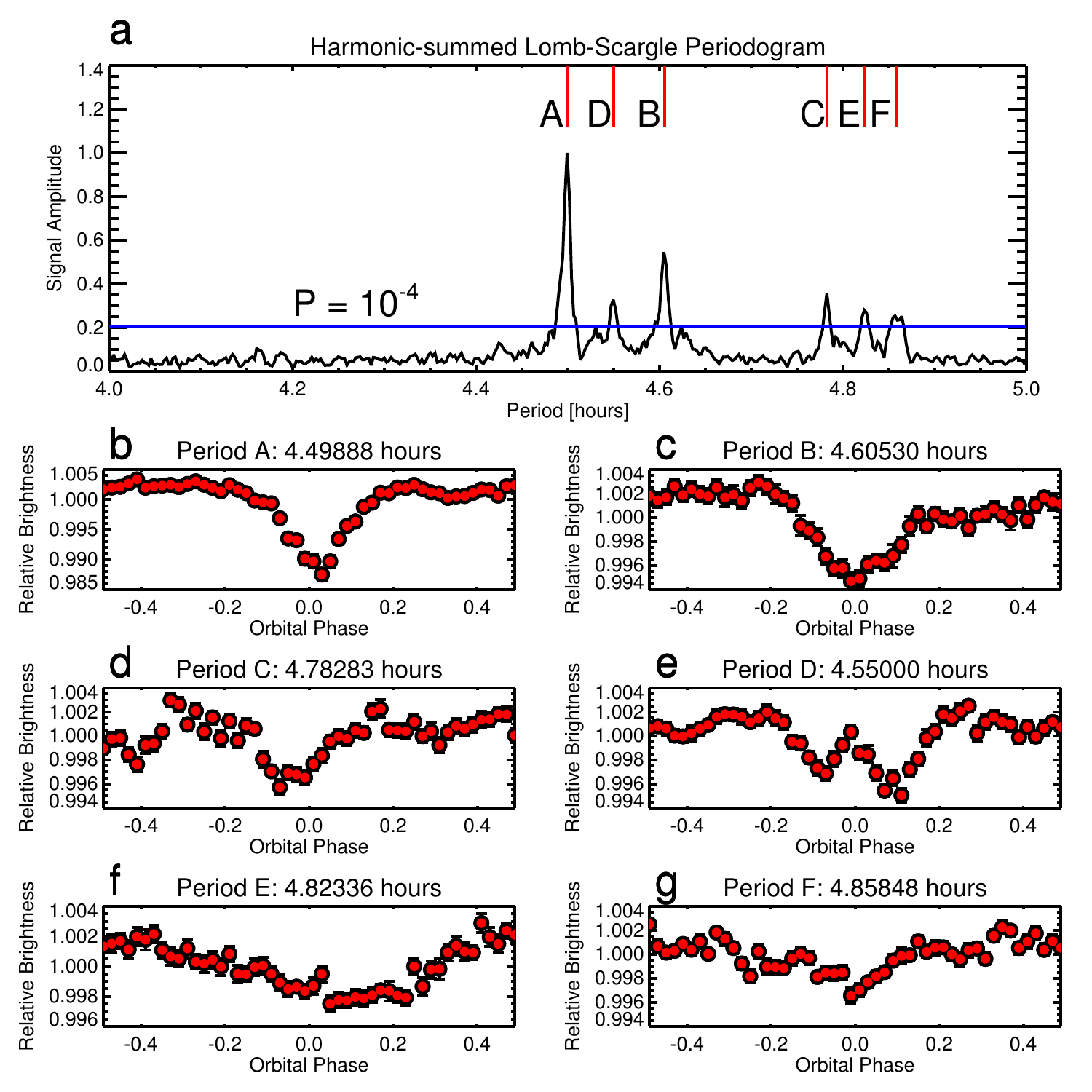}
\caption{K2 discovery observations of WD 1145+017 \citep{v15}. {\em Top panel:} Lomb-Scargle periodogram with amplitudes of the first 3 harmonics summed.  Six significant and distinct peaks are identified.  {\em Bottom panel:} Folded lightcurve for each of the detected periodicities.  The corresponding fold period is written above each panel.} 
\label{fig:fig1}
\end{figure}

A periodogram search of the K2 data revealed about 80 `ordinary' transiting planet candidates around main sequence stars, as well as a strong transit-like signal at a period of 4.5 hours from \thisstar.  The transit profile was broad (occupying nearly 30\% of an orbital cycle), had a  profile that was unlike that of a typical hard-body transit, and was variable in depth.  The periodogram and fold about the  4.4989-hour period are shown in Fig.~\ref{fig:fig1}.  A closer analysis of the K2 light curve also revealed evidence for five other significant periodicities with periods between 4.55 and 4.86 hours (designated Periods B through F), suggesting multiple bodies in close orbits.  These other periodicities can be seen in the Lomb-Scargle periodogram in Fig.~\ref{fig:fig1}, and the corresponding folded transit profiles are shown in the panels below the periodogram.

A significant limitation of the K2 data was its time sampling. For bandwidth reasons, \Kepler\ only can {\ron store} photometric data for the majority of the targets it observes, including WD 1145+017, {\ron after an integration time of 30 minutes}. Because the transit signal around \thisstar\ had a period of 4.5 hours, K2 only recorded nine photometric measurements per orbit. This limitation is particularly unfortunate for studying objects transiting or eclipsing white dwarfs, because the host star's small size means transits and eclipses happen on timescales of about 1 minute, rather than the hour to day timescales for transits of main sequence stars. Such rapid transit features are totally smeared out by the 30-minute K2 integrations. 

\begin{figure}
\centering
\includegraphics[scale=.70]{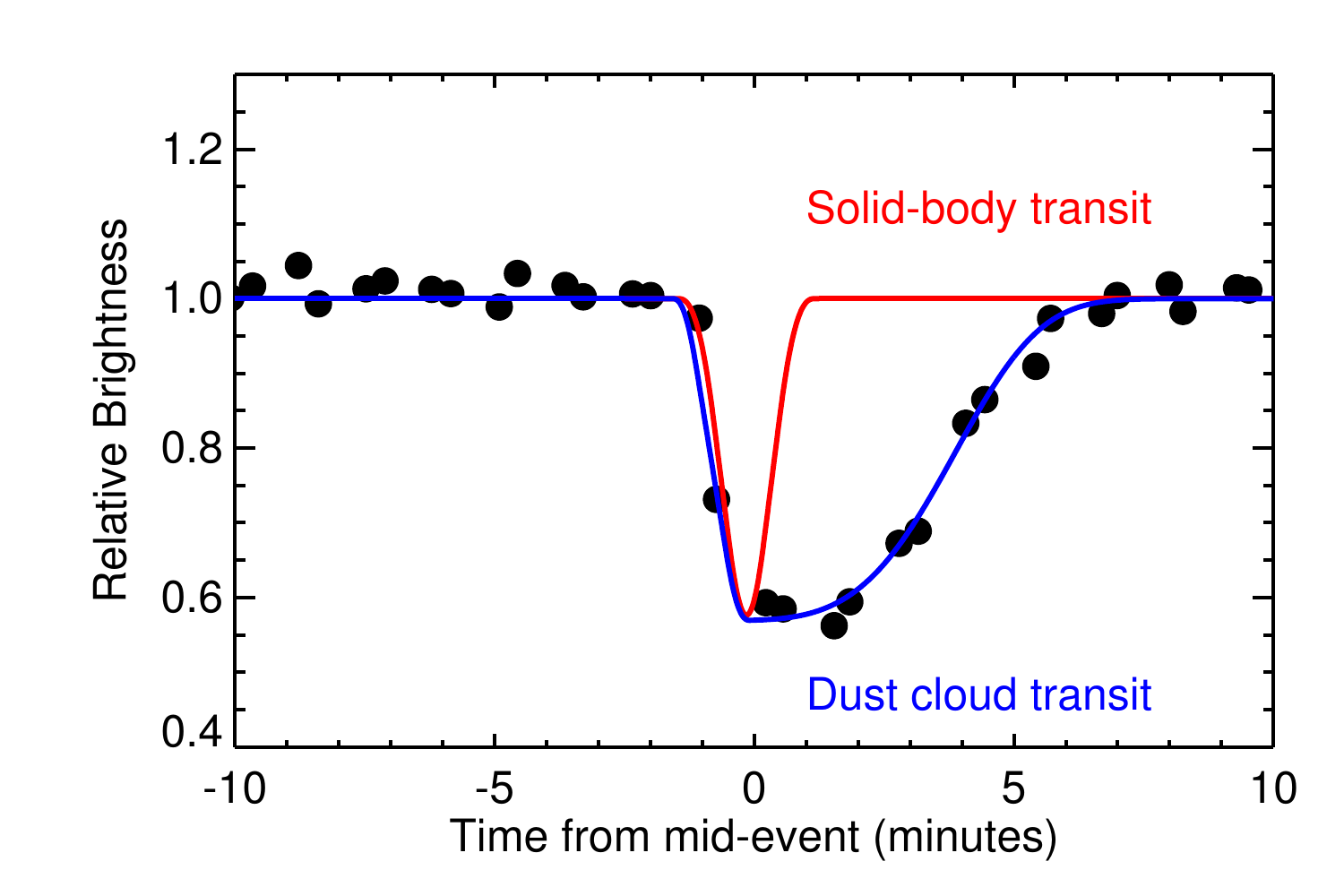}
\caption{Ground-based observations of two transits by \citet{v15}, compared to a model transit of a solid body (red) and a dust cloud (blue). {\ron We n}ote that the transit shapes do not always have the short ingress-long egress asymmetry. Follow-up observations have shown that while there is some preference for this shape, symmetric transits and long ingress-short egress asymmetries are common as well.}
\label{fig:fig2}
\end{figure}

Motivated by the very short orbital periods found in the K2 data, \citet{v15} began photometric observations of \thisstar\ using small ground-based telescopes at rapid cadence to confirm and better resolve the transits.   After a few nights of high-cadence photometric observations, several things became clear.  \thisstar\ was indeed being transited by objects in a $\simeq$4.5 hour orbital period, but the transits were not always present. The transits were deep, up to 40\% of the star's flux, and lasted about 5 minutes -- too deep and short in duration to be a transit of anything other than a white dwarf, but too long to be the transit of a small solid body across the white dwarf star (see Fig. \ref{fig:fig2}). The transits were asymmetric with a fast ingress time and slow egress time -- similar to transits of disintegrating planets observed around main sequence stars by \Kepler\ (see see van Lieshout and Rappaport, ``Disintegrating Rocky Exoplanets''; this Handbook). Finally, the transits did not always occur at the same phase of a 4.5 hour orbit. On two different nights, \citet{v15} observed two convincing transits separated by 4.5 hours, but the pairs of transits observed on each of these nights happened almost 180 degrees out of phase, with respect to a 4.5 hour orbit, further suggesting the possibility of multiple objects in orbit. 

Medium resolution spectroscopy from the MMT Observatory yielded two additional important clues. First, there was no evidence for radial velocity variations over the 4.5 hour orbital periods (with a limit of {\ron $\sim$500 m s$^{-1}$}), confirming that any bodies transiting \thisstar\ had to be of planetary mass or less. More importantly, the MMT spectrum revealed that \thisstar\ exhibited absorption lines from elements heavier than hydrogen and helium, including magnesium, aluminum, silicon, calcium, iron, and nickel. \thisstar\ is therefore a ``polluted'' white dwarf -- a member of the class of white dwarfs believed to have accreted disrupted planetary material. Data from NASA's WISE spacecraft show evidence for infrared excess emission, another hallmark of polluted white dwarfs and evidence for disrupted planetary material in orbit of the white dwarf. 

\citet{v15} interpreted the transits of \thisstar\ as being produced by dust clouds.  This inference was made by analogy with a similar phenomenon of so-called `disintegrating' planets that are found transiting main-sequence stars and which appear to exhibit dusty tails (``Disintegrating Rocky Exoplanets''; this Handbook). These objects, including KIC 12557548 \citep{kic1255}, KOI 2700 \citep{koi2700}, and K2-22 \citep{k222}, show similar features to the objects transiting \thisstar, including asymmetric transits with rapidly varying transit depths. The disintegrating planets are all in short-period orbits around their hosts stars, and are highly irradiated, to the point where rocky minerals would likely sublimate rapidly. \citet{perezbecker13} showed that a small, sub-Mercury sized object orbiting close to a host main-sequence star could undergo rapid mass loss as rocky material sublimates due to the high irradiation environment, and flows away from the planet in a Parker-type thermal wind. 

While the idea of small disintegrating rocky objects with dusty effluents causing the transits is at least partially and qualitatively successful, \citet{v15} freely admitted that the observations of \thisstar\ do not support the exact scenario seen in the systems orbiting main-sequence stars. In particular, there have been no detections of phase shifts in the transits of disintegrating planets around main-sequence stars -- when the transits of these canonical disintegrating planets appear, they always happen at the same orbital phase. The cause of the phase shifts between the different sets of transits seen by \citet{v15} was clarified by \citet{Croll15} who, acting on news of the detection of transits of \thisstar, obtained a considerably larger set of ground-based follow-up data than \citet{v15} in May 2015. Over the course of 32 hours of observations, \citet{Croll15} detected nine transits. A blind periodicity search for these transits revealed that many of the transits seemed to recur with an orbital period about 30 seconds shorter than the dominant period detected by \citet{v15} in K2 data, but that the recurring transits appeared to be scattered over a range of orbital phases. \citet{Croll15} interpreted these transits as being caused by many different discrete bodies in almost identical 4.5 hour orbits, explaining the phase shifts seen between the sets of transits detected by \citet{v15}.

\begin{figure}
\centering
\includegraphics[scale=.45]{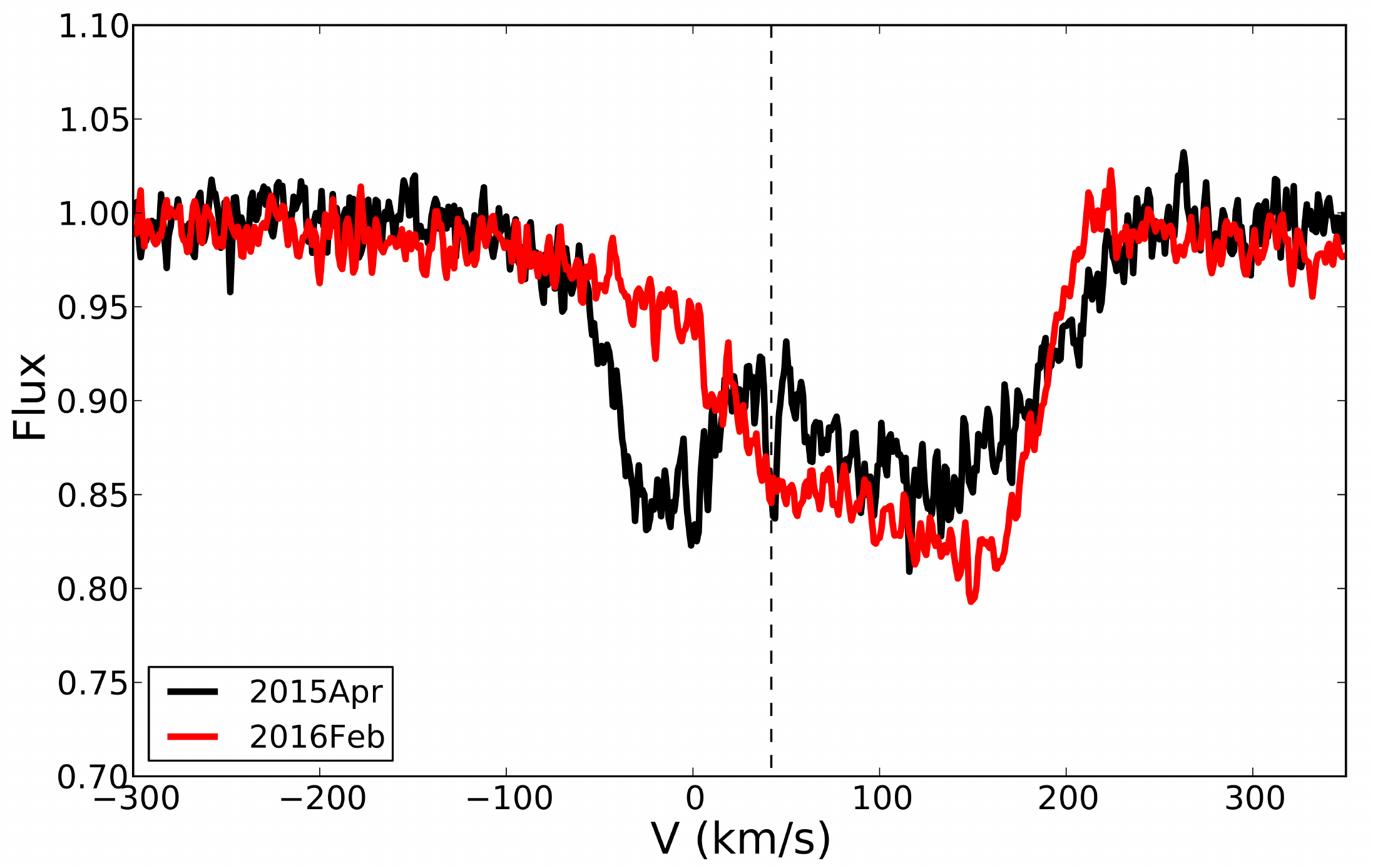}
\caption{Illustrative composite spectra (i.e., summed over 5 different metal lines) for the 2015 April observations as well as from a follow-up observation from 2016 February (Siyi Xu, 2016, private communication.).  Note the broad line widths of $\sim$300 km s$^{-1}$.  {\ron The offset of the dashed vertical line from zero indicates the gravitational redshift from the surface of the white dwarf.} }
\label{fig:fig3}
\end{figure}

Meanwhile, an independent group led by Siyi Xu also found their attention drawn to \thisstar. \citet{Xu16} were conducting a spectroscopic survey of white dwarfs with excess infrared emission at high spectral resolution with the goal of detecting heavy elements in white dwarf spectra to learn about the compositions of small extrasolar asteroids or planets. On 15 April 2015, only a few hours after \citet{v15} detected the first set of transits in their ground-based follow-up, \citet{Xu16} obtained a high signal-to-noise spectrum of WD 1145+017 with the High Resolution Echelle Spectrometer (HIRES) at Keck Observatory. As \citet{Xu16} expected from their detection of near infrared excess emission, the spectrum revealed the presence of numerous absorption features corresponding to elements like iron, silicon, nickel, magnesium, and calcium. \citet{Xu16} were surprised, however, to find broad absorption features near many of the detected metal absorption lines, presumably caused by circumstellar gas orbiting WD 1145+017.  Illustrative composite spectra (i.e., summed over 5 different metal lines) are shown in Fig.~\ref{fig:fig3} for the 2015 April observations as well as from a follow-up observation from 2016 February. The circumstellar features are broad (with line widths up to 300 \kms) and deep (obscuring up to 30\% of the star's flux at those wavelengths), and the features are visible in lines from ionized states of iron, magnesium, chromium, titanium, calcium, manganese, and nickel. 

The circumstellar absorption features found by \citet{Xu16} are unique among all other polluted white dwarfs. Some white dwarfs show circumstellar gas in emission \citep{Gansicke06, Manser16} indicating hot gas disks (not necessarily viewed edge-on), and other white dwarfs show weak circumstellar absorption features in ultraviolet spectra and optical spectra \citep{Debes12, Gansicke12}, but none shows circumstellar absorption at the strengths seen around \thisstar. The discovery of circumstellar gas absorption at \thisstar\ is independent confirmation of the presence of material orbiting close to the white dwarf between the star and our vantage point on Earth. The same favorable orbital inclination of the \thisstar\ disrupted planetary system, i.e., near $90^\circ$, is required to see both the transits and the circumstellar gas absorption.  

\section{Ground-based follow-up observations}

{\em Ground-based photometric monitoring}

\begin{figure}
\centering
\includegraphics[scale=.45]{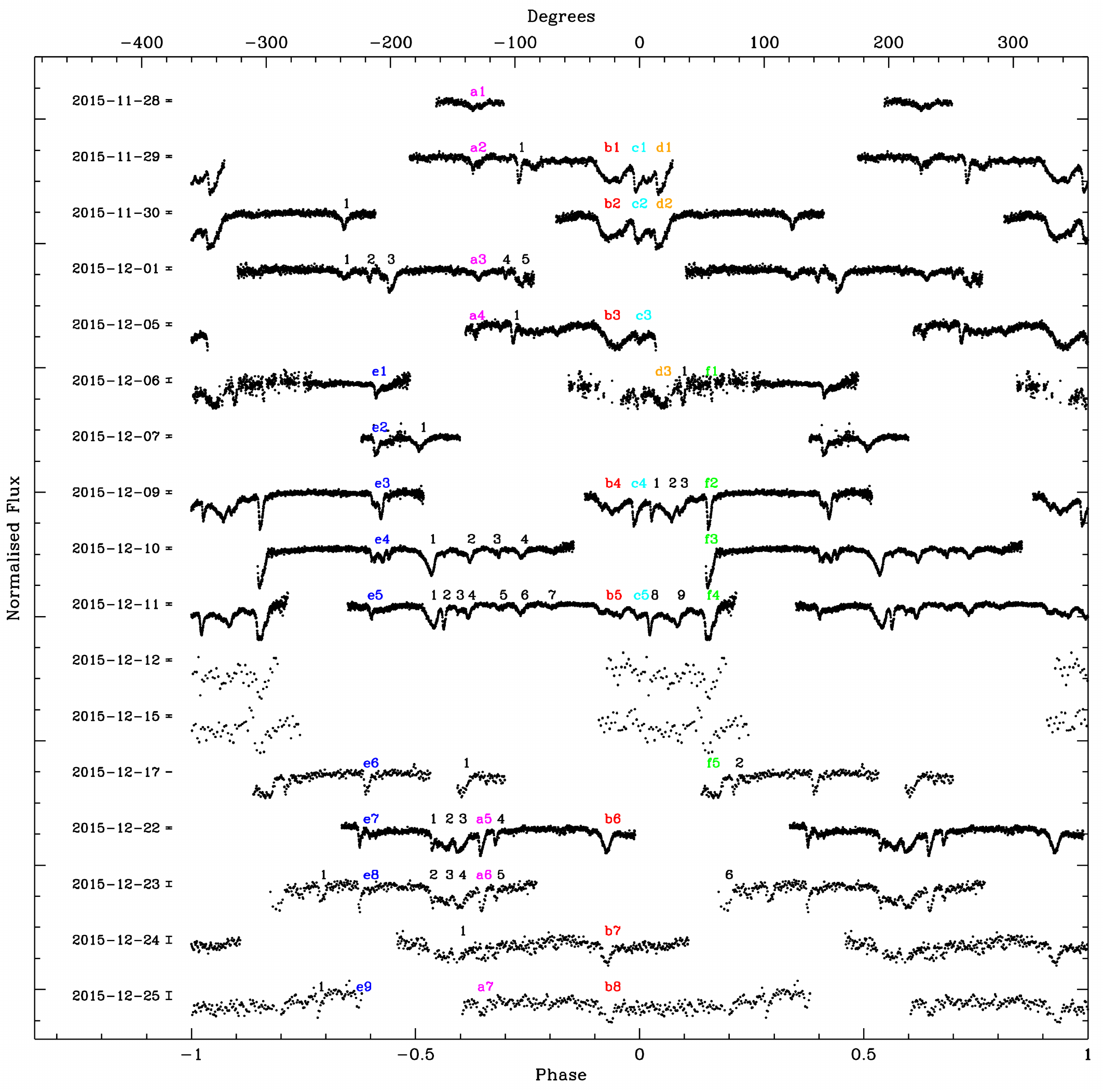}
\caption{Photometric monitoring of \thisstar\ for an interval of a month using a 2.4-m telescope (from \citealt{Gaensicke16}). The phasing of the diagram uses a period of 4.4930 hr, approximately the same as the `A fragments' detected by ground-based observations.}
\label{fig:fig4}
\end{figure}

The first extensive and systematic ground-based photometric observations of \thisstar\ commenced in 2015 November \citep{Gaensicke16,Rappaport16}.  \citet{Gaensicke16} photometrically observed \thisstar\ during 15 nights in 2015 November using 2.4\,m and 1\,m telescopes.  The results are shown in Fig.~\ref{fig:fig4}. Several distinct dips can be seen over the course of a single orbit, with some as deep as 50\%.  Averaged over the orbit, as much as $\sim$11\% of the flux is removed by the dips.  It is quite apparent from this result that the source was much more `active', in the sense of having more and deeper dips, during this period than during the K2 and the initial ground-based followup observations \citep{v15,Croll15}.  

A number of the dips seen in Fig.~\ref{fig:fig4} could be tracked from night to night, thereby allowing for more precise periods to be derived.  The periods found by \citet{Gaensicke16} range from 4.491 to 4.495 hours.  These differ by between 0.1\% and 0.2\% from the K2 `A' period of 4.4989 hours.  No sign of the K2 `B' through `F' periods was found in the \citet{Gaensicke16} data. However, if the dips at these latter periods had remained at the same depths found in the K2 observations (fractions of a percent) they {\ron could} {\em not} have been detected.  Thus, it appears that \thisstar\ was mostly active near the `A' period in 2015 November, but with much greater dip depths.

\begin{figure}
\centering
\includegraphics[scale=.35]{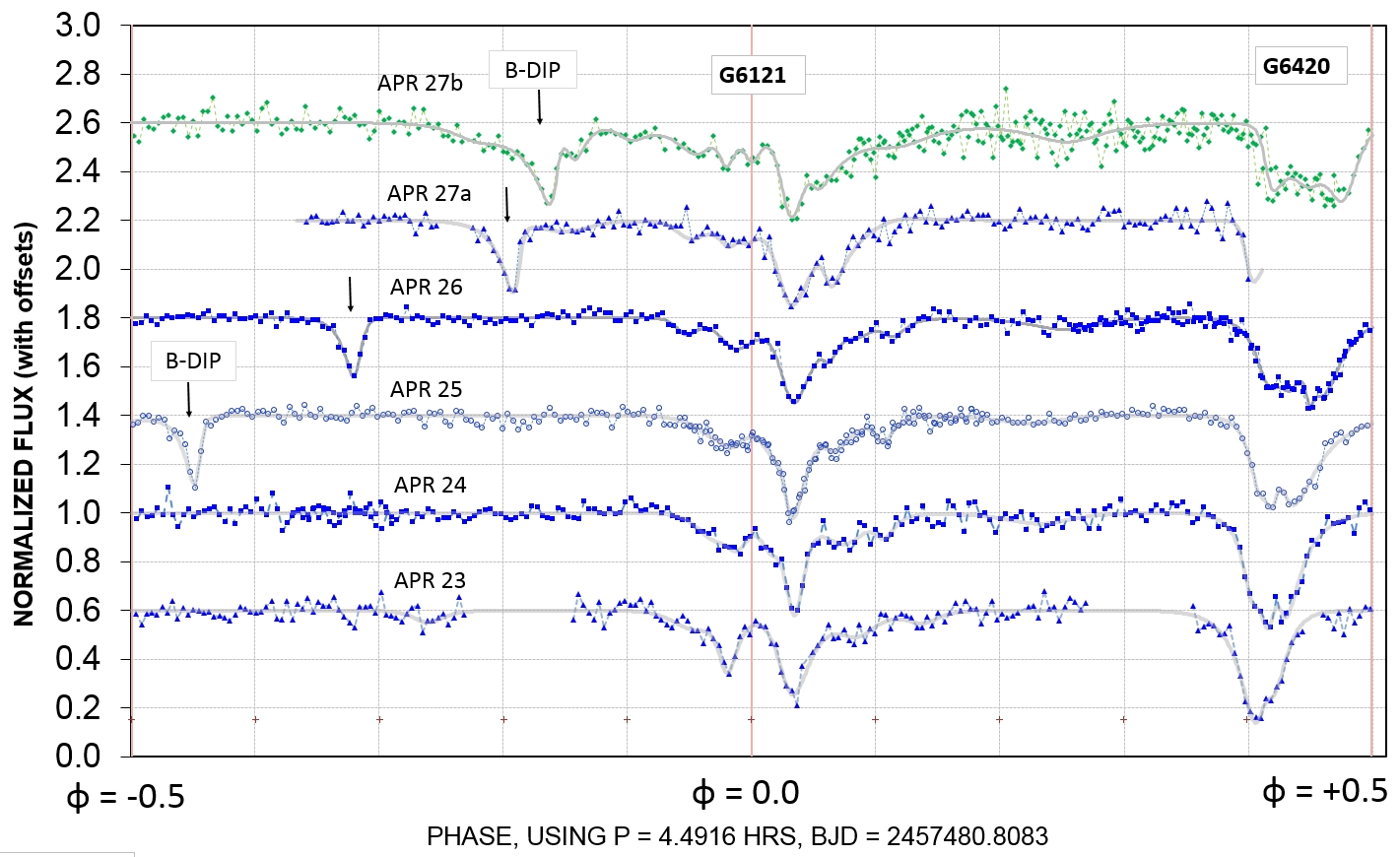}
\caption{Stack of 6 lightcurves for \thisstar\ reported by \citet{Gary16} from 2016 April.  {\ron The first 5 lightcurves (blue) are from the IAC80 32$''$ telescope and the last one (green) is from a 20$''$ telescope.}}
\label{fig:fig5}
\end{figure} 

\begin{figure}
\centering
\includegraphics[scale=.35]{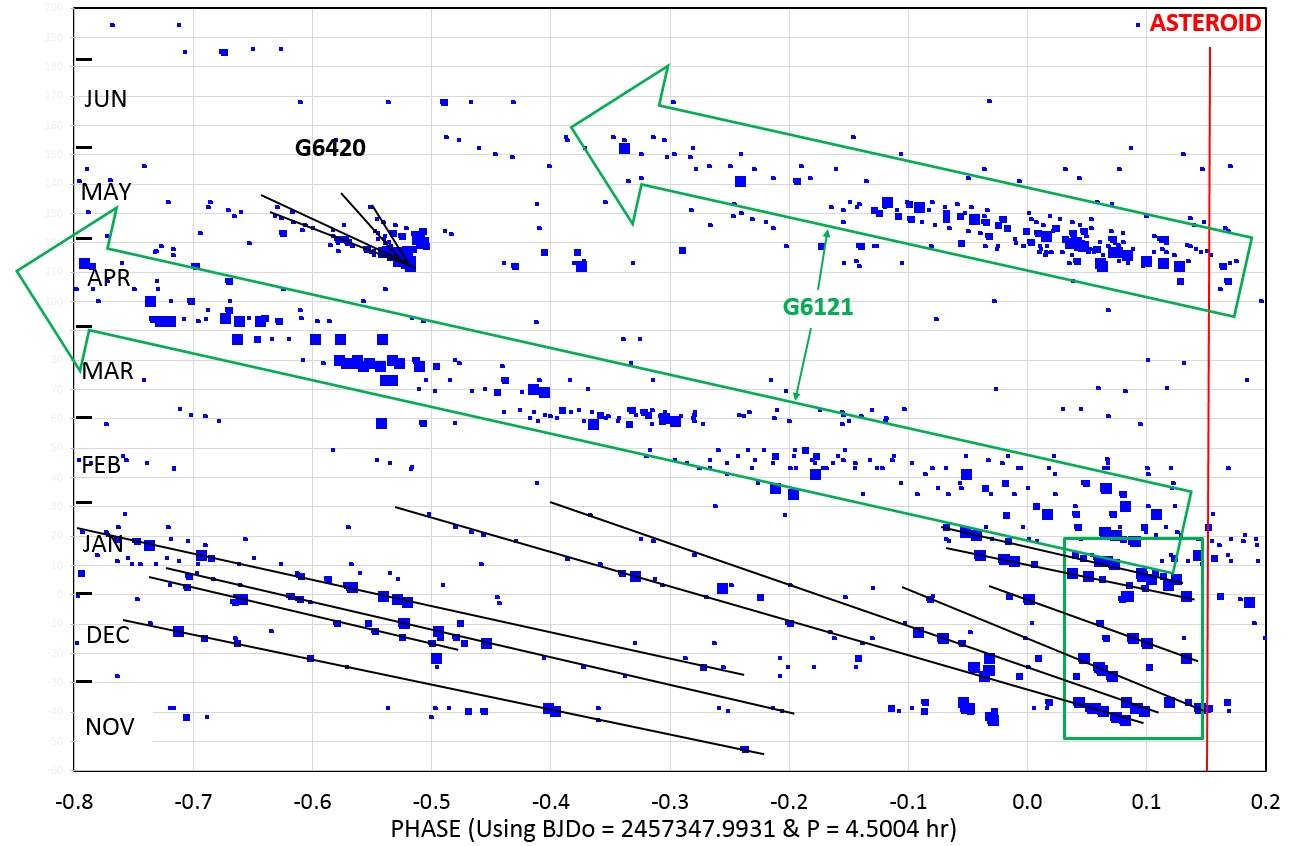}
\caption{`Waterfall' diagram for \thisstar\ of the dips recorded over 8 months from 2015 November through 2016 June \citep{Gary16}.  The phasing here is based on the `A-asteroid' period of 4.5004 hr, essentially the same as found with K2.}
\label{fig:fig6}
\end{figure} 

Starting simultaneously with the \citet{Gaensicke16} observations, \citet{Rappaport16} and \citet{Gary16} began a much longer monitoring campaign of \thisstar\ using small optical telescopes in the 30-80 cm range.  A typical set of lightcurves from {\ron five} sequential nights is shown in Fig.~\ref{fig:fig5}.  Here the data are stacked vertically and phased to an assumed period of 4.4916 hours.  The dips near phases 0 and 0.4 are nearly repeatable over the five nights.  However, there is clearly an additional dip that is moving quickly in phase, labeled `B-dip'.  This dip feature was associated by \citet{Gary16} with an object orbiting at the K2 `B' period.  

\cite{Gary16} (see also \citealt{Rappaport16}) found it useful to construct so-called `waterfall' diagrams to display the complicated temporal behavior of the dip evolution.   These diagrams are {\ron produced} as follows.  The dips are all formally fit with a simple analytic function (in this case an asymmetric hyperbolic secant, `AHS'; see \citealt{Rappaport16}), whose fitted parameters are the depth, center time, and ingress and egress times.   One can then use these parameters to {\ron plot} the kind of `waterfall' diagram shown in Fig.~\ref{fig:fig6}.  In that {\ron diagram}, each dip is represented by a rectangular bar, with orbital phase during a single night plotted in the horizontal direction, and observation night plotted vertically.  The depth of the dip is proportional to the thickness of the bar, and the duration is equal to the length of the bar. This particular plot is phased to an assumed period of 4.500 hr, the dominant mean period found by \citet{Rappaport16}.  

As one can see from Fig.~\ref{fig:fig6} there is a collection of dips between phases 0.03 and 0.15 during 2015 November and December.  After that, and for the ensuing $\sim$5 months, the orbiting objects started drifting in phase (the group labeled `G6121').  The periods associated with these drifting features in the lightcurve were in the range 4.491 to 4.495 hours, in basic accord with those found by \citet{Gaensicke16}.

\begin{figure}
\centering
\includegraphics[scale=.38]{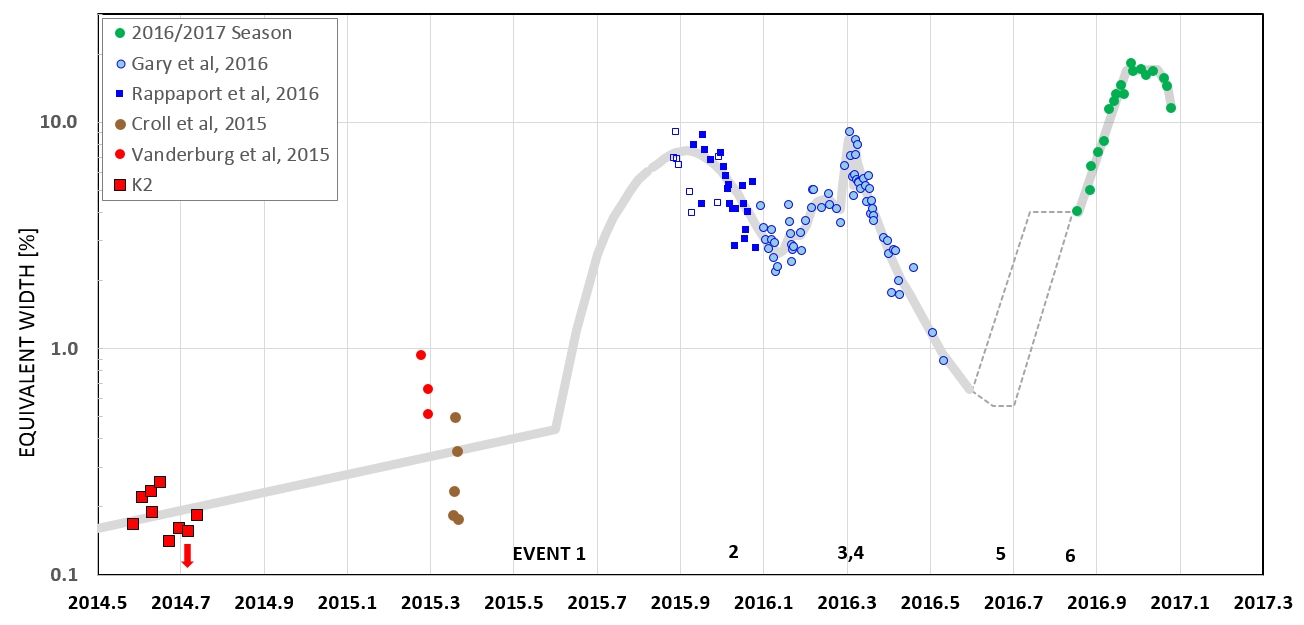} 
\caption{Activity level of \thisstar\ since its discovery with K2 (from \citealt{Gary16} and private communication).  The `activity' is defined as the dip depths averaged over an orbital cycle.}
\label{fig:fig7}
\end{figure}

We summarize here {\ron some of} what was learned after 8 months of monitoring \thisstar\ during the 2015-2016 observing season. First, the activity level of transits was quite high with dips obscuring as much as 10\% of the flux averaged around the orbit.  This is at least an order of magnitude larger than during the K2 discovery period.  The source dust activity level is shown as a function of time in Fig.~\ref{fig:fig7}.  Second, numerous different periods are seen between 4.490 to 4.500 hour, possibly indicating a dozen different bodies orbiting and emitting dusty effluents.  These are all within 0.2\% of the K2 `A' period.  Only one of the other five K2 periods, i.e., the `B' period, became active enough to detect with the small monitoring telescopes, but remained detectable for only 2-3 weeks. Third, in all photometric monitoring to date, the transits have never been deeper than 60\% of the star's total flux. Finally, we can see from Fig.~\ref{fig:fig7} that the source activity at the start of the 2016-2017 observing season is now even higher than {\ron during} all of the previous season, and has recently obscured as much as 17\% of the orbit-averaged flux, but is currently declining to lower values (B. Gary, private communication).   

{\ron Finally in regard to the `A' period, the} ground-based observations from \citet{Croll15}, \citet{Gaensicke16}, \citet{Rappaport16}, and \citet{Gary16} all find a period that is slightly, but significantly, shorter than the 4.5 hour A-period found {\ron with} K2. \citet{Rappaport16} proposed a model where a large asteroid, about 1/10 the mass of Ceres occasionally released fragments into slightly shorter period orbits.  \citet{Gurri2017} simulated the subsequent interactions of multiple fragments in such a shorter-period orbit with the parent asteroid as well as amongst themselves, while \citet{Veras17} showed explicitly how a rubble pile near the Roche limit discharges fragments into shorter (as well as longer) period orbits.  However, since early 2016, there has been no additional observational evidence that the 4.5 hour K2 A-period is persistent or ``special,'' raising some questions about this interpretation. 

\vspace{8pt}
\noindent
{\em Colour-Dependence of the Transit Depths}

	Assuming that the dips in flux from WD 1145 are a result of dusty effluents from rocky bodies, it is important to learn what we can about the nature of the dust.  Once the dust sublimates, its chemical composition, and by implication that of the planetesimal, can be investigated from its transmission spectrum as it crosses the face of the white dwarf ({\ron \citealt{Xu16}; see} the Discovery Observation section).  While the material is still in the form of dust grains, we are limited for now, to attempting to measure (i) its size distribution and (ii) some rough information about its chemical composition from both the grain scattering properties and lifetimes against sublimation. 

\begin{figure}
\centering
\includegraphics[scale=.33]{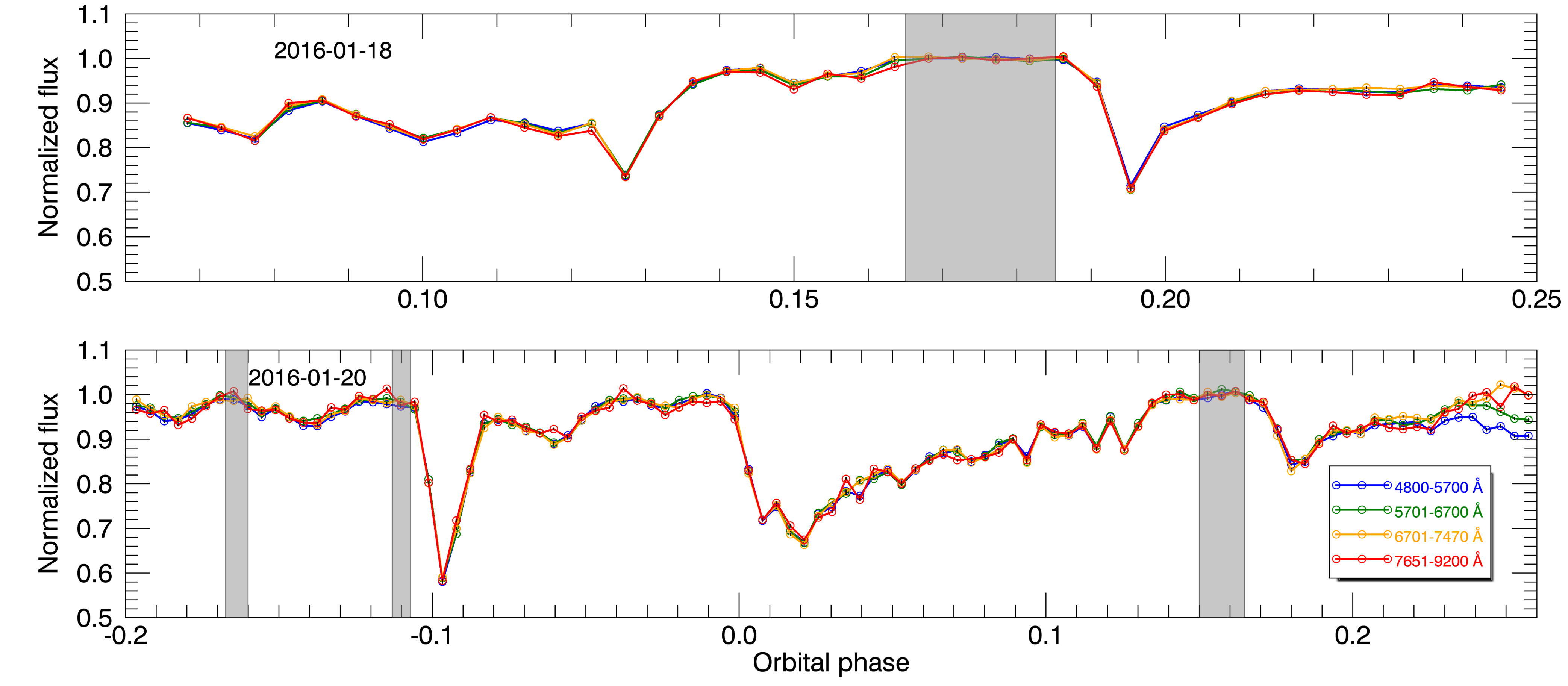}
\caption{GTC lightcurves of \thisstar\ taken simultaneously in four wavebands and covering several dips \citep{Alonso16}.  The nearly identical dip profiles in the four bands can be used to constrain the dust grain sizes to $\gtrsim 0.5 \mu$m.  The divergence of the curves after phase 0.22 in the lower panel is due to atmospheric effects.}
\label{fig:fig8}
\end{figure}

	Thus far, there have been several attempts to measure the colour-dependence of the dips in the visible (\citealt{Croll15,Alonso16}; Z.~Berta-Thompson 2016, private communication)  and in the NIR (\citealt{Zhou16}).  None of these has succeeded in finding a difference in transit depth with wavelength, and we can thereby set constraints on the dust grain sizes of $\sim$0.5 $\mu$m in the visible band and $\sim$0.8 $\mu$m in the NIR.  A good example of this type of measurement is shown in Fig.~\ref{fig:fig8}.  There are four superposed traces of the flux from WD1145 in different wavebands as a function of time over more than half an orbital cycle showing several dip features.  The four traces are in bands covering 480 nm to 920 nm, and there are no systematic differences observed among the different bands.  \citet{Alonso16} showed that the depths of these transits were the same to within a statistical precision of $\sim$1\% from which they were able to rule out small particles below $\sim$0.5 $\mu$m.  {\ron This limit is derived from the fact that for particles whose sizes are greater than the observing wavelength, the scattering cross section quickly becomes equal to the geometric area of the dust grain.  When that occurs, the scattering becomes independent of wavelength.}

\vspace{8pt}
\noindent
{\em Time-resolved spectroscopy}

Continued observations of the circumstellar gas absorption lines (\citealt{Redfield16}; S. Xu 2016, private communication) show that the shape and intensity of these broad lines are highly variable on different timescales.  One possible explanation for the broad and shifted absorption lines is a mildly eccentric ring of gas orbiting close to the white dwarf (see also \citealt{Redfield16}).  It can be shown that for a mildly eccentric orbit (i.e., with $e \lesssim 0.1$), the radial velocity of the gas as a function of its transverse location, $x$, across the face of the host star, is given by:
\begin{equation}
v_r \simeq v_0 \left(e \cos \omega - x/a\right) ~\simeq ~ \frac{2030}{P_{\rm min}^{1/3}} \, e \cos \omega - \frac{900}{P_{\rm min}} \left(\frac{x}{R_{\rm wd}}\right) ~~{\rm km~s}^{-1}
\end{equation}
where $\omega$, $v_0$, and $a$ are the {\ron argument} of periastron, orbital speed, and semimajor axis for the orbiting ring of gas, respectively.  The expression on the right is specifically for the WD 1145+017 system, $P_{\rm min}$ is the orbital period expressed in units of minutes, and $R_{\rm wd}$ is the radius of the white dwarf.  Thus, for example, if a line is observed with a range of wavelengths corresponding to radial velocities of $-100$ to +200 km s$^{-1}$, this yields $e \cos \omega \simeq 0.045$ and {\ron $P \simeq 6$ minutes}.

\section{Some Theoretical Considerations}

\noindent
The pollution of white dwarfs had been a long-standing problem in astrophysics, dating back to the 1970s \citep{fontaine79}, when it was realized that in order for heavy elements to appear in the spectra of white dwarfs, they must have been recently accreted from external sources (or they would have quickly sunk to the center of the white dwarfs, where they would be un-observable). The history of this problem is extensively discussed in ``Characterizing Planetary Systems Around White Dwarfs'' (B.~Zuckerman and E. Young; this Handbook).  In brief, the discovery of dusty debris disks near many polluted white dwarfs \citep{Zuckerman87, Becklin05} and the fact that the abundances of heavy elements found in white dwarfs were quite similar to the elemental ratios in rocky solar system bodies \citep{Zuckerman06} led astronomers to believe that these heavy elements were likely the disrupted remains of rocky objects from the white dwarf progenitors' planetary systems.

\vspace{8pt}
\noindent
{\ron {\em Origin of the Debris:} }

\citet{Debes02} were among the first to suggest that heavy element pollution in white dwarfs was caused by the remnants of ancient planetary systems. Using analytic arguments and numerical studies, \citet{Debes02} showed that when an evolving star undergoes mass loss, previously stable planetary systems could become unstable to close encounters. The critical Hill separation for two equal-mass planets, $\Delta_c$, in terms of the semi major axes of two planets, $a_1$ and $a_2$, is given by: 
\begin{equation}
\Delta_c  = \frac{a_2-a_1}{a_1} \simeq \sqrt{\frac{8}{3}(e_1^2 + e_2^2) + 9 \left( \frac{m}{M}\right)^{2/3}} 
\end{equation}
where $e_1$ and $e_2$ are the eccentricities of the two planets, $m$ is the mass of each of the two planets, and $M$ is the mass of the star. \citet{Debes02} noticed that when the stellar mass $M$ decreases (for a white dwarf progenitor, often by a factor of two or more), the critical separation increases, which can push previously stable systems into the unstable regime, rearranging the planetary system dynamically. This rearrangement, \citet{Debes02} speculated, could cause an increased rate of comets from the outer parts of the planetary system to be scattered inwards to close encounters with the white dwarf. 

\begin{figure}
\centering
\includegraphics[scale=.90]{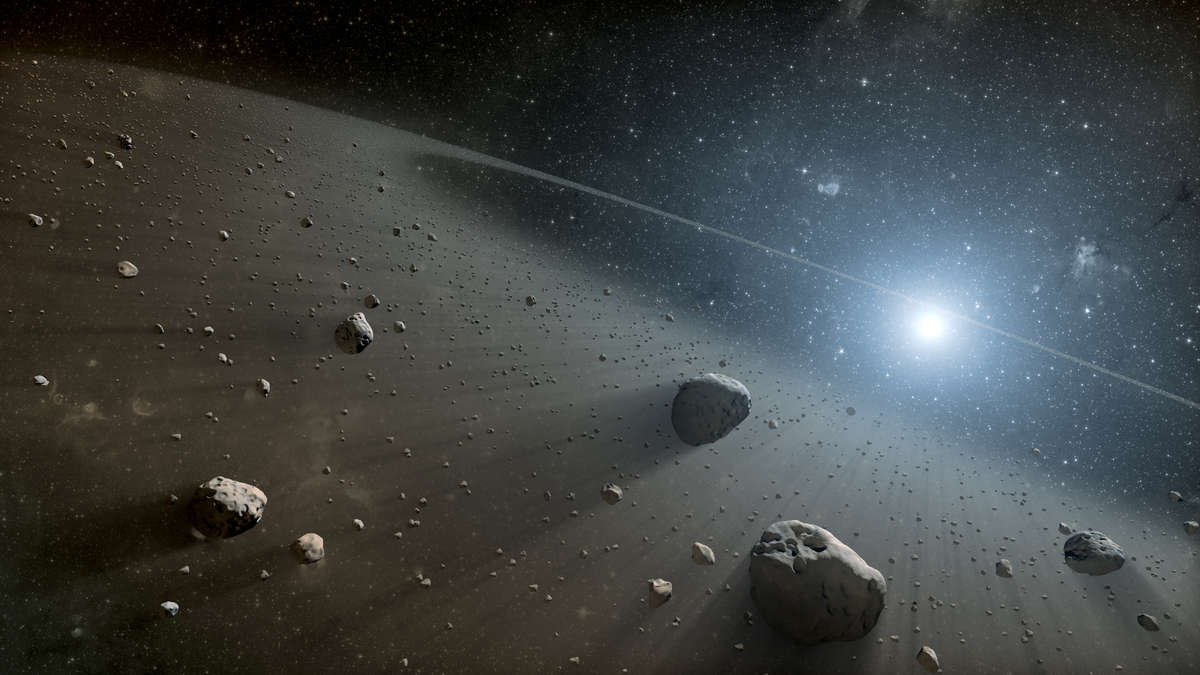}
\caption{Artist conception of what the \thisstar\ debris disk system might look like as viewed from one of the orbiting asteroids. {\ron Image credit: NASA/JPL-Caltech} }
\label{fig:fig9}
\end{figure} 

Since then, other mechanisms have been proposed to bring planetary material to close encounters with the host white dwarfs. \citet{Debes2012b} showed that stellar mass loss can sweep asteroids into mean motion resonances with giant planets {\ron (akin to Jupiter)}, where the asteroids are perturbed into highly eccentric orbits and close encounters with the white dwarf. Similarly, \citet{Bonsor2011} showed that stellar mass loss in a planetary system can perturb objects from an exo-Kuiper belt close to the white dwarf. Additionally, numerical studies have shown that for systems that border on Hill instability, increasing the Hill radii of the planets can in many cases cause the planetary systems themselves to become unstable, and undergo strong dynamical interactions on relatively short timescales, leading to close periastron passages {\ron with} the white dwarf \citep{Veras2013, Veras2016c}.

\vspace{8pt}
\noindent
{\ron {\em Roche Limit:}} 

After rocky bodies have been dynamically perturbed too close to the host white dwarf, they are subject to tidal breakup.  The famous result of Edouard Roche, cast in the context of bodies orbiting \thisstar\, can be written as:
\begin{equation}
s_{\rm crit} = \xi R_* \left(\rho_*/\rho_p\right)^{1/3}
\label{eqn:Roche}
\end{equation} 
where $s_{\rm crit}$ is the critical distance from a star of density $\rho_*$ that a planetesimal of uniform density $\rho_p$ can come before it is tidally split.  For Roche's original problem he found that the constant $\xi$ had a value of 2.45.  Newer treatments of the `tidal splitting' problem have been carried out since then, and many of these are summarized by {\ron \citeauthor{Davidsson} (1999;} see also \citealt{Holsapple06}).  Taking into account such issues as the criteria for actually splitting the body into separated pieces; what trajectory the body is on; whether the body is corotating with its orbit; and the role that material strength plays, $\xi$ is found to lie closer to the range of $1.2 \lesssim \xi \lesssim 1.7$.  There is also the issue of how centrally concentrated the mass of the planetesimal is, e.g., the case where its central density may be considerably higher than in its mantle.  In the limiting (unrealistic) case where the planetesimal can be considered to have all its mass concentrated at the center, one can use the Roche potential to show that in this case $\xi \simeq 2.1$.  

All of this pertains to cases where the planetesimal under question is larger than a few km in size.  Otherwise, for small bodies, they may simply be held together by solid-state forces amounting to tensile strengths that are large compared to the pressures imposed by gravity near the center of the body.

One can rewrite Eqn.~(\ref{eqn:Roche}) in a more convenient way for our problem by cubing both sides 
and solving for the critical density (i.e., below which the body is subject to tidal breakup), to find:
\begin{equation}
\rho_{\rm crit}  = \frac{3 \xi^3 M_*}{4 \pi s_{\rm crit}^3} = \frac{3 \pi  \xi^3}{G} \frac{1}{P_{\rm orb}^2} \simeq 87 \,(\xi/2)^3 \frac{1}{P_{\rm hr}^2} ~{\rm g~cm}^{-3}
\end{equation}
where in the second term on the right we made use of Kepler's 3rd law, and where in the final expression, $\xi$ has been normalized to a typical value of 2.  For a period of 4.5 hours, as in the case of the innermost objects orbiting around \thisstar\, we find critical densities of 7.9, 5.0, 2.6, and 1 {\ron g cm$^{-3}$}, depending on whether one uses Roche's value of $\xi$, the value of $\xi$ from the Roche potential, or other more recent values for $\xi$, respectively.

\vspace{8pt}
\noindent
{\ron {\em Debris Rings:} }

Once large rocky bodies have been tidally disrupted due to their close approaches to the white dwarf host, the result will be an eccentric ring of large chunks of debris \citep{Veras14, Veras15}. Through a process known as a ``collisional cascade''  (see, e.g., \citealt{Kenyon02}; \citealt{Kenyon04}; \citealt{Kenyon16}; \citealt{Wyatt11}) the sizable rocky bodies in this orbiting ring of debris undergo a sequence of collisions which eventually break down the few large objects into a very wide array of object sizes.  These will range from planetesimals, to asteroids, rocks, pebbles, and dust, with a rough power-law size distribution.  

In Figure \ref{fig:fig9} we show an artist's conception of what such a debris disk might look like.  Note the size distribution ranging from asteroids down to dust.

\begin{figure}
\centering
\includegraphics[scale=.75]{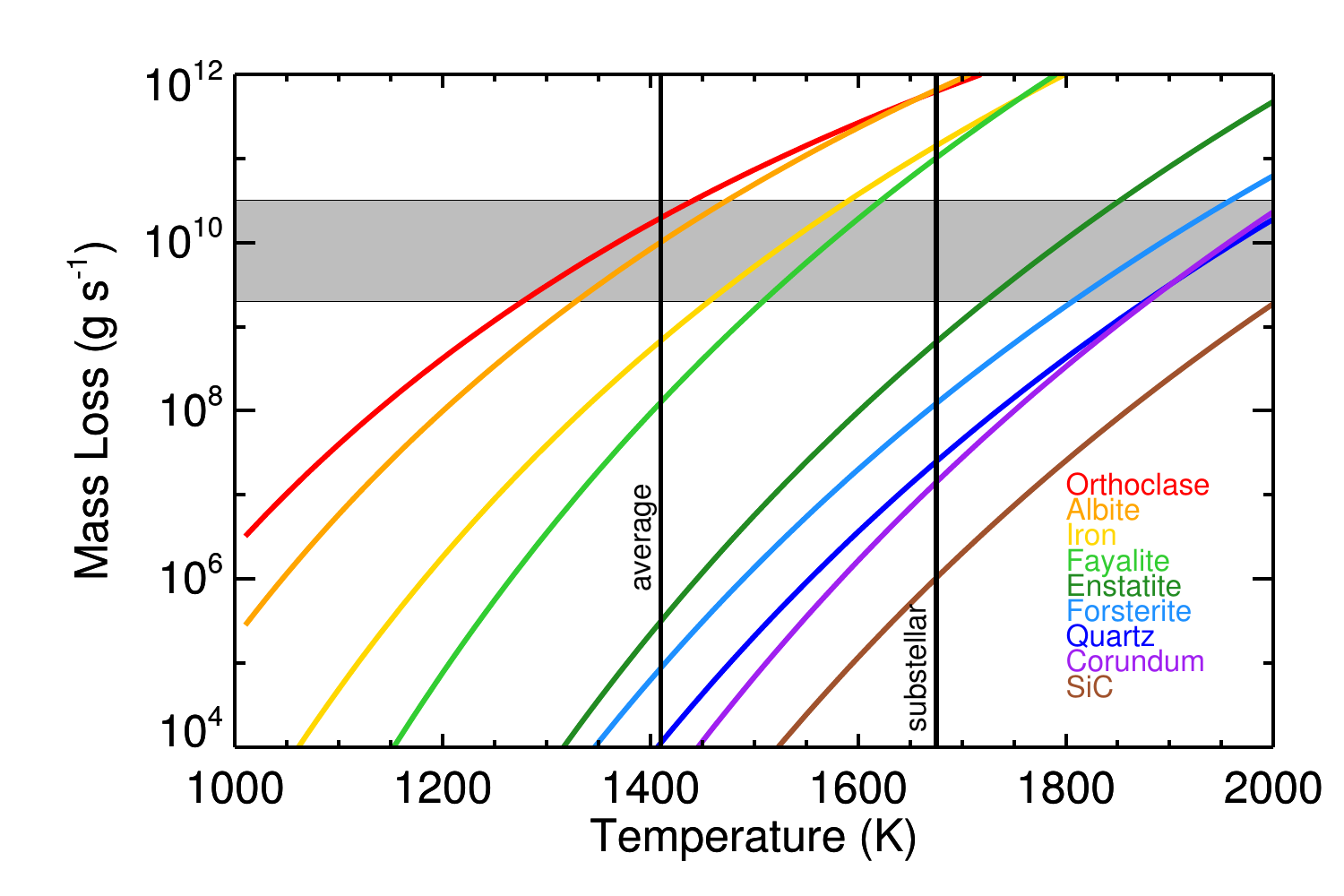}
\caption{Sublimation mass-loss rates for different minerals on the surface of a Ceres-sized body orbiting \thisstar\ (from \citealt{v15}).}
\label{fig:fig10}
\end{figure} 

At the current epoch we are apparently witnessing the debris disk in WD 1145+017 after it has undergone a collisional cascade -- which may, in fact, still be ongoing. The main goal of any comprehensive model would be to relate the properties of such a debris disk to the observational dipping activity in WD 1145+017, the high-velocity gas absorption lines, and the excess NIR emission.   The asymmetric transit profile and wildly variable transit depths observed {\ron from} \thisstar\ are qualitatively similar to transits of disintegrating planets around main sequence stars (``Disintegrating Rocky Exoplanets''; this Handbook), and it is believed that similar physical processes may be responsible for both types of transits, but with some significant differences. Perhaps the primary difference between the dusty-tailed planets around main-sequence stars and the situation in WD 1145+017 is the existence of numerous orbiting bodies in the white-dwarf system as opposed to a single more-substantive body orbiting the main-sequence stars.  As we have seen, there are likely a dozen or more independently orbiting bodies in WD 1145+017 of a size that can emit sufficient dust to block a significant fraction of the white dwarf's light.

\section{Explaining the Transits}

The transits around \thisstar\ set it apart among all other polluted white dwarfs, and understanding what causes them is crucial to understanding the system as a whole. Among the various dips in flux, a number of them are uniquely identifiable, long lasting (i.e., for up to months), and of more or less constant shape and duration. Here, we describe three possible models of how the starlight is blocked. 

\vspace{8pt}
\noindent {\em Continuous Dust Production Sublimating from Small Rocky Bodies}

The first model for explaining the transits of \thisstar\ is that dust is emitted continually from an orbiting rocky body at a relatively constant rate -- merely giving the illusion of a permanent feature. This scenario requires a substantive underlying body in a fairly stable orbit to produce the deep and persistent transits observed. For tidally locked objects in nearly circular 4.5 hour orbits around \thisstar, the temperature of the object at the substellar point is about 1675 K and the average temperature over the dayside hemisphere is a bit lower at about 1410 K. These temperatures are high enough that many minerals will sublimate. The vapor pressure, $p_{\rm vap}$, 
and the knowledge that any vapors produced will freely stream away from the surface of such low-mass objects, allow us to calculate a mass loss flux $J$ from the surface of the body into a vacuum: 
\begin{equation}\label{masslossflux}
J = \alpha~p_{\rm vap}\sqrt{\frac{\mathcal{M}}{2\pi k_{B}T_{\rm eq}}}  ~~~~~{\rm where}~~~~~p_{\rm vap} = \exp \left[-\frac{\mathcal{M} L_{\rm sub}}{k_{B}T_{\rm eq}}+b\right]
\end{equation}
and where $\mathcal{M}$ is the material's molecular mass, $L_{\rm sub}$ is the material's latent heat of sublimation, $k_{B}$ is Boltzmann's constant, $T_{\rm eq}$ is the temperature of the body, $b$ is an empirically measured constant, and $\alpha$ is the ``sticking coefficient'' and is roughly 0.1 to 0.3. \citet{v15} calculated mass loss rates for various different materials from a single Ceres sized object orbiting \thisstar\ and found that several different materials and minerals, in particular orthoclase, albite, iron, and fayalite could sublimate and produce dust at the rates necessary to explain the transits of \thisstar.  We show in Fig.~\ref{fig:fig10} mass loss rates due to sublimation for a Ceres size object as a function of the surface equilibrium temperature for a variety of common minerals. Figure 8 of van Lieshout and Rappaport (``Disintegrating Rocky Exoplanets''; this Handbook) gives the sublimation rate as a function of planetesimal size for a fixed surface temperature of 2000 K.

The realization from more recent large photometric monitoring campaigns  \citep{Gaensicke16, Gary16} that there are likely many smaller fragments producing deep transits and that these fragments are likely considerably smaller in order to coexist in {\ron very close orbits} \citep{Rappaport16} makes it more difficult to account for the deep transits via sublimation of refractory minerals. \citet{v15} calculated that the transits seen around \thisstar\ require mass-loss rates of order 10$^9$ g\,s$^{-1}$, and found that the mass loss fluxes calculated from Equation \ref{masslossflux} could reproduce those mass loss rates if the surface area was that of a Ceres sized object with a radius of roughly 500 km. However, if the fragments producing most of the transits {\ron at} the A-period are actually closer in size to Halley's comet ($\sim$5-10 km radius), as suggested by \citet{Rappaport16}, then their surface areas are roughly 10$^3$-10$^4$ times smaller than \citet{v15} assumed. At the same mass loss fluxes, it therefore becomes difficult (but not out of the question) for sublimation of refractory elements to eject enough material to produce the observed transits. Some scenarios which might mitigate this problem and yield increased mass loss from small fragments include the presence of volatile elements (like water ice), which sublimates much faster than refractory materials like iron, or a higher surface area to mass ratio for the fragments than previously assumed. Alternatively, mass loss may not be completely driven by sublimation, and other processes such as volcanic activity or even collisions may contribute to the ejection of material causing the deep transits we see. 

\vspace{8pt}
\noindent {\em Persistent dust from impulsive collisions}

The second scenario involves an impulsive ejection of dust.  In this scenario, dust is ejected in discrete events. The wavelength dependence of the Mie absorption cross sections (and the corresponding inferred emissivities) for smaller dust particles causes these particles to be quickly heated to quite high temperatures (i.e., $\gtrsim 2000 $ K) which leads to rapid sublimation.  The result may be that the surviving grains are only those with sizes $\gtrsim$ few $\mu$m.  The calculated values of $\beta$, the ratio of radiation pressure forces to gravity, for these larger residual grains may be as low as $\beta \simeq 0.001$. Such small values of $\beta$ result in minimal radiation pressure to push the grains into a different orbit than the emitter, and, likewise, the orbital decay timescale due to the Poynting-Roberston effect may be much longer than the observation timescale.  Therefore, in this scenario, the dust is released and simply hangs around the emitting asteroid for considerable intervals of time.  The main effect in spreading such a dust cloud may be shearing due to its finite radial extent.  If the cloud is `ribbon-like', i.e., narrow in the radial direction, then it can avoid shearing.  However, if it is extended by as much in the radial direction as it is perpendicular to the mid-plane (i.e., comparable to the size of the white dwarf), then the cloud would shear completely around the orbit within a couple of days.

\vspace{8pt}
\noindent {\em Azimuthal Asymmetry in a Dust Ring}

A final scenario invokes a quasi-permanent dust ring orbiting \thisstar\ viewed edge on, which has azimuthal density asymmetries that cause differential extinction. When an over-dense region passes in front of the host star, it would cause a transit-like event. These asymmetries could not be self-gravitating -- no reasonable masses \citep{Gurri2017, Veras2016b} could create a Hill sphere large enough to produce the dips we see -- so the asymmetries would likely have to be driven by some orbiting shepherding objects, such as for some of the structures found in Saturn's rings \citep[see e.g.,][]{Murray05,Murray08,Hyodo15}. There are several observational arguments against this scenario.  The first is that we often see transits, or diminishings of brightness, but not brightenings of the star.  One would expect that the presence of over-densities that cause dimming events would also be accompanied by the presence of under-densities, which would cause brightening events when they pass in front of the host star. Second, the density waves in such a dust ring caused by shepherding bodies would be expected to show dip structures that have more manifest symmetries, and would not likely produce the isolated sharp narrow dip features that are sometimes observed in the lightcurves.  Finally, it is not obvious how such an obscuring dust ring would change in overall density by more than an order of magnitude over the course of months as is seen (e.g., Fig.~\ref{fig:fig7}).

\section{Future Work} 

The \thisstar\ system is highly complex, rapidly evolving, and poses difficult observational and theoretical challenges. However, there are several paths forward to learning more about this object and the general process of how planetary material is tidally disrupted. In this section, we discuss both observational and theoretical studies which should be undertaken to further our understanding.  

\vspace{0.1cm}
\noindent {\em Continued Photometric Monitoring:} 
The simplest future observations of \thisstar\ are perhaps among the most important. In particular, it is crucial to continue observing \thisstar\ photometrically with small to moderate-class telescopes. Continued photometric monitoring will allow us to track the transit activity level and determine the timescales on which the transit activity level changes. These observations are setting the first observational constraints on how long the tidal disruption of rocky bodies around white dwarfs last and how they progress, which are leading to new insights about the nature of the disrupted object \citep{Veras17}. A more ambitious and far-reaching goal for photometric monitoring would be to detect and characterize objects orbiting in the sub-dominant B-F periods in the same way that objects in the A-period have been characterized to date. Now that photometric monitoring has revealed an object transiting at the sub-dominant B-period from K2, it is reasonable to believe that continued monitoring may reveal the frequency of transits of the B-F periods found in K2 relative to the numerous transits associated with the A-period, and lead to a stronger understanding of why the A-period experiences higher activity than the other periods. Finally, it would be good to carry out at least some photometry with large-aperture telescopes at shorter time-cadence to look for low-level and rapid timescale variability.

\vspace{0.1cm}
\noindent {\em Continued High-Resolution Spectroscopic Monitoring:}
High resolution spectroscopy of \thisstar\ has revealed intriguing patterns and changes in the circumstellar absorption features \citep{Xu16, Redfield16}, but these observations are in general photon-starved. At 17th magnitude, high spectral resolution observations of \thisstar\ are just barely possible on the short ($\lesssim$ 5 minute) timescales of the photometric variability with 10-meter class optical telescopes today. Future observations of \thisstar\ with the next generation of efficient high-resolution spectrographs on 30-meter class telescopes should yield significantly superior signal-to-noise ratios on the timescales of transits, making a detailed examination of the short-timescale variability possible and revealing the interplay between the transiting dust and circumstellar gas. 

\vspace{0.1cm} 
\noindent {\em UV spectroscopy:}
Ultraviolet spectroscopy of polluted white dwarfs often reveals the presence of additional elements in their spectra, and it is likely that \thisstar\ will be no exception. UV spectroscopy of \thisstar\ is crucial to obtain while the {\em Hubble Space Telescope} (HST) is still operating in order to learn more about the composition of the disintegrating material and the circumstellar gas \citep[which has previously been found in other white dwarfs in the ultraviolet, ][]{Gansicke12}. These observations are forthcoming; an ongoing HST program led by Siyi Xu is obtaining ultraviolet spectroscopy of \thisstar\ with the Cosmic Origins Spectograph (COS). 

\vspace{0.1cm}
\noindent {\em Gaia Parallax:}
ESA's Gaia mission will measure parallaxes of over a billion stars, including WD 1145+017. By comparing the precise trigonometric distance to \thisstar\ with a photometric distance estimate, it will be possible to determine whether \thisstar\ is significantly extincted, and therefore to infer the presence of dust in the line of sight (even while out of transit). 

\vspace{0.1cm}
\noindent {\em Mid-Infrared Observations:}
Multi-wavelength observations of debris disks around white dwarfs can yield information about the extent and orientation of the disks \citep{Jura03}. This information is especially interesting in the case of \thisstar\ because of the additional information we know about the system, in particular the orbital radii and inclination of the transiting debris. Naively, one would expect that for an edge-on system with an optically thick debris disk, where the orbits and debris disk are viewed at an inclination angle close to $90^\circ$, the luminosity of the disk would be relatively small, yet for the edge-on orbits at \thisstar\, the disk luminosity is somewhat large. 

Multi-wavelength photometric observations in the mid-infrared can yield information about the orientation and size of the debris disk that are not possible with current data (which only extends out to about 5 microns). Since the end of the \Spitzer\ cryogenic mission in 2009, there are no telescope resources capable of detecting such faint sources at wavelengths beyond 5 microns, but the {\em James Webb Space Telescope}, scheduled to launch in 2018, will once again provide that capability at unprecedented sensitivity. Measuring the \thisstar\ debris disk's spectral energy distribution into the mid-infrared could indicate a misalignment between the debris disk and the transiting material's orbits, or it could show that the disk is likely not optically thin \citep[like the disk surrounding SDSS J155720.77+091624.6][]{Farihi17}. 

{\ron Finally, with regard to mid-IR observations, it is worth noting that measurements of the silicate feature at 10 $\mu$m can directly reveal the presence of circumstellar silicates in dust grains (see, e.g., \citealt{Jura09}).}

\vspace{0.1cm}
\noindent {\em Polarimetry:} 
 It would be quite challenging to measure possible polarization of light from WD 1145+017.  Because there is not likely to be any preferred grain alignment via magnetic fields, any polarization would be due to scattering.  If polarization can be measured, this would be quite useful in confirming the presence of dust and learning more about its spatial distribution.

\vspace{0.1cm}
\noindent {\em {\ron Questions for Future Theoretical and Observational} Investigations:}
Finally, here we list {\ron five} questions which we believe are necessary to answer in the near future if we are to fully understand the \thisstar\ system. While continued and further observations will help, these questions will likely require detailed calculations and study. \\

\noindent
$\bullet$ Is the dust production that feeds any of the dust clouds continuous, {\ron i.e., due to sublimation}, or episodic, e.g., due to collisions of rocky bodies? \\
$\bullet$ If the dust production is episodic, then what keeps dust-cloud dip features orbiting coherently, in some cases for months?  The dust clouds must be considerably larger than the Hill's sphere of any plausible body that is emitting the dust, and therefore the dust must be unbound.   \\
$\bullet$ What causes the dramatic changes in dust cloud numbers and optical depth on timescales of days to years, like the change between observations by \citet{v15} and \citet{Croll15}, and those by \citet{Gaensicke16} and \citet{Gary16} a year later? \\
$\bullet$ What sets the apparent maximum flux decrease of $\sim$60\% in the light of the white dwarf?  Is this a result of changes in the optical depth of dust that covers essentially all of the white dwarf surface, or due to optically thick clouds that can, for some reason, never cover more than 60\% of the white dwarf surface? \\
$\bullet$  {\ron How do the dust clouds become large enough to cause the large transit depths we see? In order to obscure large portions of the white dwarf the debris must be orbiting in different planes that are tilted by a range of angles spanning at least {\ron $\pm 1/2^\circ$}. In turn, this implies velocity components that are perpendicular to the mean orbital plane of $\sim$3 km s$^{-1}$, which are much too large for thermal speeds.  This seems to require collisions.  However, such high speeds would lead to rapid angular spreading of any resultant dust clouds that is much larger than observed. Finite sublimation lifetimes may help in regard to this latter issue.}

{\ron Future observations and inquiries along these lines} should improve our understanding of the many processes at work around polluted white dwarfs like \thisstar\ and contribute to our knowledge of the ultimate fate of planetary systems (including our own solar system) after their host stars retire from the main sequence.

\begin{acknowledgement}
The authors are grateful to Bruce Gary, Siyi Xu, and Zach Berta-Thompson for quite helpful discussions and for sharing some of their unpublished results regarding WD 1145+017.  {\ron We also acknowledge Ben Zuckerman for his insightful comments about the manuscript.}
\end{acknowledgement}

\end{document}